\documentclass[12pt]{article}

\usepackage{anysize}
\marginsize{2cm}{2cm}{2cm}{3cm}

\usepackage{times,amssymb,mathptm,graphicx}

\newcommand{\ket}[1]{|#1\rangle}

\newcommand{\thm}[3]{
  \begin{#1}
    #2
    \\
    {\em {\bf Proof:}
    #3}
    \hfill \fbox{}
  \end{#1}
}

\newcommand{\thmx}[2]{
  \begin{#1}
    #2
  \end{#1}
}

\title{
  Synthesis of Reversible Logic Circuits
  \thanks{
    \footnotesize
    This work was partially supported by
    the Undergraduate Summer Research Program
    at the University of Michigan and by the DARPA QuIST program.
    The views and conclusions contained herein are those of the authors
    and should not be interpreted as necessarily representing official
    policies or endorsements, either expressed or implied, of the
    Defense Advanced Research Projects Agency (DARPA) or the U.S. Government.
  }
}

\author{
  Vivek V. Shende \\  {\tt vshende} \and
  Aditya K. Prasad \\  {\tt akprasad} \and
  Igor L. Markov \\ {\tt imarkov} \and
  John P. Hayes \\ \tt{jhayes} \and
  Advanced Computer Architecture Laboratory {\tt @umich.edu} \and
  University of Michigan, Ann Arbor, MI 48109-2122\\
}

\date{}

\begin{document}

\maketitle

\vspace{20mm}

\abstract{ \addtolength{\baselineskip}{1mm}
  Reversible or information-lossless circuits have
  applications in di\-gital signal processing, communication,
  computer graphics and cryptography. They are also a fundamental requirement
  in the emerging field of quantum computation. We investigate the synthesis of
  reversible circuits that employ a minimum number of gates
  and contain no redundant input-output line-pairs (temporary
  storage channels). We prove constructively that
  every even permutation can be
  implemented without temporary storage using NOT, CNOT and TOFFOLI
  gates. We describe an algorithm for the synthesis of optimal circuits
  and study the reversible functions on three wires,
  reporting the distribution of circuit sizes.
  Finally, in an application important to quantum computing, we synthesize
  oracle circuits for Grover's search algorithm, and show a significant
  improvement over a previously proposed synthesis algorithm.
}

\newpage

\section{Introduction}
\label{sec:intro}

In most computing tasks, the number of output bits is relatively
small compared to the number of input bits. For example, in a
decision problem, the output is only one bit (yes or no) and the
input can be as large as desired. However, computational tasks in
digital signal processing, communication, computer graphics, and
cryptography require that all of the information encoded in the
input be preserved in the output. Some of those tasks are
important enough to justify adding new microprocessor instructions
to the HP PA-RISC (MAX and MAX-2), Sun SPARC (VIS), PowerPC
(AltiVec), IA-32 and IA-64 (MMX) instruction sets
\cite{RubyLee,RubyLee2}. In particular, new bit-permutation
instructions were shown to vastly improve performance of several
standard algorithms, including matrix transposition and DES, as
well as two recent cryptographic algorithms Twofish and Serpent
\cite{RubyLee2}. Bit permutations are a special case of {\em
reversible functions}, that is, functions that permute the set of
possible input values. For example, the butterfly operation
$(x,y)\rightarrow (x+y,x-y)$ is reversible but is not a bit
permutation. It is a key element of Fast Fourier Transform
algorithms and has been used in application-specific Xtensa
processors from Tensilica. One might expect to get further
speed-ups by adding instructions to allow computation of an
arbitrary reversible function. The problem of chaining such
instructions together provides one motivation for studying
reversible computation and reversible logic circuits, that is,
logic circuits composed of gates computing reversible functions.

Reversible circuits are also interesting because the loss of information
associated with irreversibility implies energy loss \cite{Bennett}.
Younis and Knight
\cite{YounisK94} showed that some reversible circuits can be made
asymptotically energy-lossless as their delay is allowed to grow
arbitrarily large. Currently, energy losses due to irreversibility
are dwarfed by the overall power dissipation, but this may change
if power dissipation improves. In particular, reversibility is
important for nanotechnologies where switching devices with gain
are difficult to build.

Finally, reversible circuits can be viewed as a special case of
quantum circuits because quantum evolution must be reversible
\cite{NielsenChuang}. Classical (non-quantum) reversible gates are
subject to the same ``circuit rules,'' whether they operate on
classical bits or quantum states. In fact, popular universal gate
libraries for quantum computation often contain as subsets
universal gate libraries for classical reversible computation.
While the speed-ups which make quantum computing attractive are
not available without purely quantum gates, logic synthesis for
classical reversible circuits is a first step toward synthesis of
quantum circuits. Moreover, algorithms for quantum communications
and cryptography often do not have classical counterparts because
they act on quantum states, even if their action in a given
computational basis corresponds to classical reversible functions
on bit-strings. Another connection between classical and quantum
computing comes from Grover's quantum search algorithm
\cite{Grover}. Circuits for Grover's algorithm contain large parts
consisting of NOT, CNOT and TOFFOLI gates only
\cite{NielsenChuang}.

We review existing work on classical reversible circuits. Toffoli
\cite{Toffoli} gives constructions for an arbitrary reversible or
irreversible function in terms of a certain gate library. However,
his method makes use of a large number of temporary storage
channels, i.e. input-output wire-pairs other than those on which
the function is computed (also known as {\em ancilla bits}). Sasao
and Kinoshita show that any conservative function ($f(x)$ is
conservative if $x$ and $f(x)$ always contain the same number of
1s in their binary expansions) has an implementation with only
three temporary storage channels using a certain fixed library of
conservative gates, although no explicit construction is given
\cite{Sasao}. Kerntopf uses exhaustive search methods to examine
small-scale synthesis problems and related theoretical questions
about reversible circuit synthesis \cite{Kerntopf}. There has also
been much recent work on synthesizing reversible circuits that
implement non-reversible Boolean functions on some of their
outputs, with the goal of providing the quantum phase shift
operators needed by Grover's quantum search algorithm
\cite{Japanese, Koreans, Travaglione}. Some work on local
optimization of such circuits via equivalences has also been done
\cite{Koreans, Japanese}. In a different direction, group theory
has recently been employed as a tool to analyze reversible logic
gates \cite{Storme} and investigate generators of the group of
reversible gates \cite{DeVos}.

Our work pursues synthesis of optimal reversible circuits which
can be implemented without temporary storage channels. In Section
\ref{sec:theory}, we show by explicit construction that any
reversible function which performs an even permutation on the
input values can be synthesized using the CNTS (CNOT, NOT,
TOFFOLI, and SWAP) gate library and no temporary storage. An
arbitrary (possibly odd) permutation requires at most one channel
of temporary storage for implementation. By examining circuit
equivalences among generalized CNOT gates, we derive a canonical
form for CNT-circuits. In Section \ref{sec:algos} we present
synthesis algorithms for implementing any reversible function by
an optimal circuit with gates from an arbitrary gate library.
Besides branch-and-bound, we use a dynamic programming technique
that exploits reversibility. While we use gate count as our cost
function throughout, this method allows for many different cost
functions to be used. Applications to quantum computing are
examined in Section \ref{sec:grover}.

\section{Background}
\label{sec:background}

In conventional (irreversible) circuit synthesis, one typically
starts with a universal gate library and some specification of a
Boolean function. The goal is to find a logic circuit that
implements the Boolean function and minimizes a given cost metric,
e.g., the number of gates or the circuit depth. At a high level,
reversible circuit synthesis is just a special case in which no
fanout is allowed and all gates must be reversible.

\subsection{Reversible Gates and Circuits}

\thmx{DEF}{ \label{def:revgate}
  A gate is reversible if the (Boolean) function it
  computes is bijective.
}

If arbitrary signals are allowed on the inputs, a necessary condition
for reversibility is that the gate have the same number of
input and output wires. If it has $k$ input and output wires,
it is called a $k\times k$
gate, or a gate on $k$ wires. We will think of the $m$th input
wire and the $m$th output wire as really being the same wire. Many
gates satisfying these conditions have been examined in the literature
\cite{PQLG}. We will
consider a specific set defined by Toffoli \cite{Toffoli}.

\thmx{DEF}{ \label{def:cnot}
  A $k$-CNOT is a $(k+1)\times (k+1)$ gate. It leaves the first $k$
  inputs unchanged, and inverts the last iff all others are $1$.
  The unchanged lines are referred to as {\em control lines}.
}

Clearly the $k$-CNOT gates are all reversible. The first three of
these have special names. The $0$-CNOT is just an inverter or NOT
gate, and is denoted by N. It performs the operation $(x)\to
(x\oplus 1)$, where $\oplus$ denotes XOR. The $1$-CNOT, which
performs the operation $(y,x)\to(y,x\oplus y)$ is referred to as a
Controlled-NOT \cite{Feynman}, or CNOT (C). The $2$-CNOT is
normally called a TOFFOLI (T) gate, and performs the operation
$(z,y,x)\to(z,y,x\oplus yz)$. We will also be using another
reversible gate, called the SWAP (S) gate. It is a $2 \times 2$
gate which exchanges the inputs; that is, $(x,y)\to (y,x)$. One
reason for choosing these particular gates is that they appear
often in the quantum computing context, where no physical
``wires'' exist, and swapping two values requires non-trivial
effort. \cite{NielsenChuang}. We will be working with circuits
from a given, limited-gate library. Usually, this will be the CNTS
gate library, consisting of the CNOT, NOT, and TOFFOLI, and SWAP
gates.

\thmx{DEF}{ \label{def:revcirc}
  A well-formed reversible logic circuit is an acyc\-lic
  combinational logic circuit
  in which all gates are reversible, and are interconnected without fanout.
}

\begin{figure}[b]
  \begin{center}
    \includegraphics[height = 1.5cm]{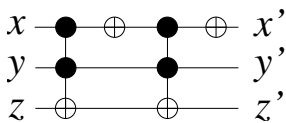}
  \end{center}
  \caption{$3 \times 3$ reversible circuit with two T gates and two N
  gates. \label{fig:nctpic}}
\end{figure}

As with reversible gates, a reversible circuit has the same number
of input and output wires; again we will call a reversible circuit
with $n$ inputs an $n\times n$ circuit, or a circuit on $n$ wires.
We draw reversible circuits as arrays of horizontal lines
representing wires. Gates are represented by vertically-oriented
symbols. For example, in Figure \ref{fig:nctpic}, we see a
reversible circuit drawn in the notation introduced by Feynman
\cite{Feynman}. The $\oplus$ symbols represent inverters and the
$\bullet$ symbols represent controls. A vertical line connecting a
control to an inverter means that the inverter is only applied if
the wire on which the control is set carries a $1$ signal. Thus,
the gates used are, from left to right, TOFFOLI, NOT, TOFFOLI, and
NOT.

\begin{figure}
  \small
  \begin{center}
      $
      \begin{array}{|ccc|ccc|}
        \hline
        x & y & z & x'& y' & z' \\ \hline
        0 & 0 & 0 & 0 & 0 & 0 \\
        0 & 0 & 1 & 0 & 0 & 1 \\
        0 & 1 & 0 & 0 & 1 & 1 \\
        0 & 1 & 1 & 0 & 1 & 0 \\
        1 & 0 & 0 & 1 & 0 & 0 \\
        1 & 0 & 1 & 1 & 0 & 1 \\
        1 & 1 & 0 & 1 & 1 & 1 \\
        1 & 1 & 1 & 1 & 1 & 0 \\ \hline
      \end{array}
      $
    \caption{
      Truth table for the circuit in Figure \ref{fig:nctpic}.
      \label{fig:nctpic:tt}
    }
  \end{center}
\end{figure}

Since we will be dealing only with bijective functions, i.e.,
permutations, we represent them using the {\em cycle notation}
where a permutation is represented by disjoint cycles of
variables. For example, the truth table in Figure
\ref{fig:nctpic:tt} is represented by $(2,3)(6,7)$ because the
corresponding function swaps 010 (2) and 011 (3), and 110 (6) and
111 (7). The set of all permutations of $n$ indices is denoted
$S_n$, so the set of bijective functions with $n$ binary inputs is
$S_{2^n}$. We will call $(2,3)(6,7)$ {\em CNT-constructible} since
it can be computed by a circuit with gates from the CNT gate
library. More generally:

\begin{figure}[b]
  \begin{center}
    \includegraphics[height=2cm]{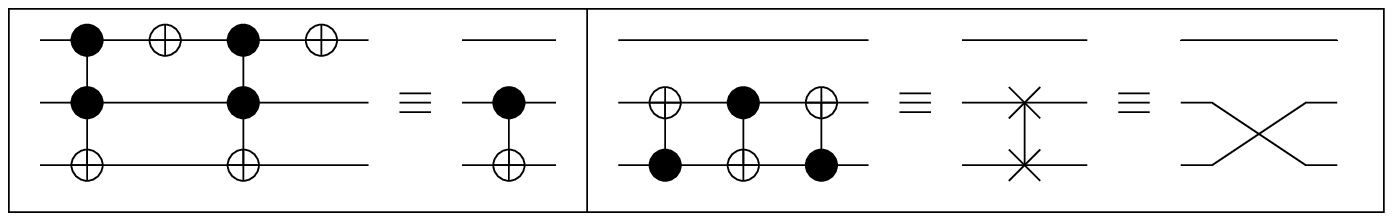} \\
    (a) \hspace{57mm} (b)
  \end{center}

  \caption{
    \label{fig:moreeq} Reversible circuit equivalences: (a)
   $T^3_{1,2} \cdot N^1 \cdot T^3_{1,2} \cdot N^1 = C^3_2$,
    (b) $C^2_3 \cdot C^3_2 \cdot C^2_3= S^{2,3}$;
    subscripts identify ``control bits'' while
    superscripts identify bits whose values actually change.
  }

\end{figure}

\thmx{DEF}{ \label{def:constr}
  Let $L$ be a (reversible) gate library. An
  $L$-circuit is a circuit composed only of gates from $L$.
  A permutation $\pi \in S_{2^n}$ is $L$-constructible if it
  can be computed by an $n\times n$ $L$-circuit.
}

Figure \ref{fig:moreeq}$a$ indicates that the circuit in Figure
\ref{fig:nctpic}$a$ is equivalent to one consisting of a single C
gate. Pairs of circuits computing the same function are very
useful, since we can substitute one for the other. On the right,
we see similarly that three C gates can be used to replace the S
gate appearing in the middle circuit of Figure
\ref{fig:moreeq}$b$. If allowed by the physical implementation,
the S gate may itself be replaced with a wire swap. This, however,
is not possible in some forms of quantum computation
\cite{NielsenChuang}.  Figure \ref{fig:moreeq} therefore shows us
that the C and S gates in the CNTS gate library can be removed
without losing computational power. We will still use the CNTS
gate library in synthesis to reduce gate counts and potentially
speed up synthesis. This is motivated by Figure \ref{fig:moreeq},
which shows how to replace four gates with one C gate, and thus up
to 12 gates with one S gate.

\begin{figure}[t]
  \begin{center}
    \includegraphics[height=3cm]{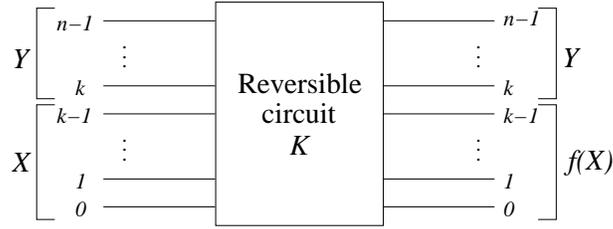}
  \end{center}
  \vspace{-5mm}
  \caption{
    \label{fig:tempstor}
    Circuit $C$ with $n-k$ wires $Y$ of temporary storage.
  }
\end{figure}

Figure \ref{fig:tempstor} illustrates the meaning of ``temporary
storage'' \cite{Toffoli}. The top $n-k$ lines transfer $n-k$
signals, collectively designated $Y$, to the corresponding wires
on the other side of the circuit. The signals $Y$ are arbitrary,
in the sense that the circuit $K$ must assume nothing about them
to make its computation. Therefore, the output on the bottom $k$
wires must be only a function of their input values $X$ and not
of the ``ancilla'' bits $Y$, hence the bottom output
is denoted $f(X)$. While the
signals $Y$ must leave the circuit holding the same values they
entered it with, their values may be changed during the
computation as long as they are restored by the end. These wires
usually serve as an essential workspace for computing $f(X)$. An
example of this can be found in Figure \ref{fig:moreeq}$a$: the C
gate on the right needs two wires, but if we simulate it with two
N gates and two T gates, we need a third wire. The signal applied
to the top wire emerges unaltered.

\thmx{DEF}{
  Let $L$ be a reversible gate library. Then $L$ is
  universal if for all $k$ and all permutations $\pi \in S_{2^k}$,
  there exists some $l$
  such that some $L$-constructible circuit computes $\pi$ using $l$ wires
  of temporary storage.
}

The concept of universality differs in the reversible and
irreversible cases in two important ways. First, we do not allow
ourselves access to constant signals during the computation, and
second, we synthesize whole permutations rather than just
functions with one output bit.

\subsection{Prior Work}

It is a result of Toffoli's that the CNT gate library is
universal; he also showed that one can bound the amount of
temporary storage required to compute a permutation in $S_{2^n}$
by $n-3$. Indeed, much of the reversible and quantum circuit
literature allows the presence of polynomially many temporary
storage bits for circuit synthesis. Given that qubits are a
severely limited resource in current implementation technologies,
this may not be a realistic assumption. We are therefore
interested in trying to synthesize permutations using no extra
storage. To illustrate the limitations this puts on the set of
computable permutations, suppose we restrict ourselves to the C
gate library. The following results are well-known in the quantum
circuits literature \cite{PQLG, QuantAlg}. We provide proofs both
for completeness, and to accustom the reader to techniques we will
require later.

\thmx{DEF}{A function $f:\{0,1\}^n \to \{0,1\}^m$ is linear iff
  $f(\mathbf{x} \oplus \mathbf{y})=f(\mathbf{x})\oplus f(\mathbf{y})$,
  where $\oplus$ denotes bitwise XOR.
}

This is just the usual definition of linearity where we think of $\{0,1\}^n$
as a vector space over the two-element field $\mathbb{F}_2$.
In our work $n=m$ because of reversibility. Thus, $f$ can be thought of
as a square matrix over  $\mathbb{F}_2$.
The composition of two linear functions is a linear function.

\thm{LEM}{ \label{lem:constr:c} \cite{QuantAlg}
  Every C-constructible permutation computes an invertible
  linear transformation. Moreover, every invertible linear transformation
  is computable by a C-constructible circuit. No C-circuit
  requires more than $n^2$ gates.
} {
  To show that all C-circuits are
  linear, it suffices to prove that each C gate computes a linear
  transformation. Indeed, $C(x_1\oplus y_1, x_2 \oplus y_2)= (x_1
  \oplus y_1, x_1 \oplus y_1 \oplus x_2 \oplus y_2) = (x_1, x_1
  \oplus y_1) \oplus (x_2, x_2 \oplus y_2) = C(x_1,y_1) \oplus
  C(x_2,y_2)$. In the basis
  $10\ldots 0$, $01 \ldots 0$, $\ldots$, $0 \ldots 01$, a C gate with
  the control on the $i$-th wire and the inverter on the $j$-th applied
  to an arbitrary vector will add the $i$-th entry to the $j$-th. Thus,
  the matrices corresponding to individual C gates account for all the
  elementary row-addition matrices.
  Any invertible matrix in $GL(\mathbb{F}_2)$ can
  be written as a product of these. Thus, any invertible linear
  transformation can be computed by a C-circuit. Finally, any
  matrix over $\mathbb{F}_2$ may be row-reduced to
  the identity using fewer than $n^2$ row
  operations.
}

One might ask how inefficient the row reduction algorithm is in
synthesizing C-circuits. A counting argument can
be used to find asymptotic lower bounds on the longest circuits
\cite{Silke}.

\thm{LEM}{\label{lem:gatebound}
  Let $L$ be a gate library; let $K_n \subset S_{2^n}$ be the
  set of $L$-constructible permutations on $n$ wires, and let
  $k_i$ be the cardinality of $K_i$. Then the longest gate-minimal
  $L$-circuit on $n$ wires has more than $\log k_n/\log b$
  gates, where $b$ is the number of one-gate circuits on $n$ wires.
  $b = poly(n)$, so for large $n$, worst-case circuits have
  length $\Omega(\log k_n / \log n)$. } {
  Suppose the longest gate-minimal $L$-circuit has $x-1$ gates.
  Then every permutation in $K_n$ is computed by an $L$-circuit
  of at most $x-1$ gates. The number of such circuits is
  $\sum_{i=1}^{x-1} b^i = < b^x$.
  Therefore, $k_n < b^x$, and it follows that $x >
  \log k_n / \log b$.

  Finally, let G be a gate in $L$ with the largest number of inputs, say
  $p$. Then, on $n$ wires, there are at most $n(n-1)\ldots (n-p+1) <
  n^p$ ways to make a $1$-gate circuit using G. If $L$ has $q$ gates in total,
  then $b \le q n^p = poly(n)$.
  Hence, $x > \log k_n / ( p \log n + \log q ) = \Omega (\log k_n / \log n)$.
}

\begin{figure}[t]
  \begin{center}
    \includegraphics[height=1.2cm]{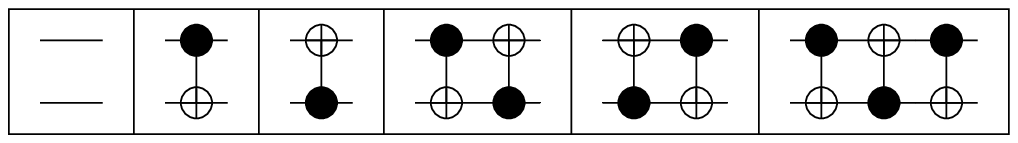}
  \end{center}
  \vspace{-4mm}
  \caption{
    \label{fig:cnots}
    Optimal C-circuits for C-constructible permutations on 2 wires.
  }
\end{figure}

We now need to count the number of C-constructible permutations. On two wires,
there are six, corresponding to the six circuits in Figure \ref{fig:cnots}.

\thm{COR}{ \cite{Silke} \label{cor:cnotcount}
 $S_{2^n}$ has $\prod_{i=0}^{n-1} (2^n - 2^i)$
 C-constructible permutations. Therefore, worst-case C-circuits
 require $\Omega(n^2/\log n)$ gates.
} {
  A linear mapping is fully defined by its values on basis
  vectors. There are $2^n-1$ ways of mapping the $2^n$-bit
  string $10...0$. Once we have fixed its image, there are $2^n-2$
  ways of mapping $010...0$, and so on.  Each basis bit-string
  cannot map to the subspace spanned by the previous bit-strings.
  There are $2^n-2^i$ choices for the $i$-th basis bit-string.
  Once all basis bit-strings are mapped, the mapping of the rest
  is specified by linearity.
  The number of C-constructible permutations on $n$ wires is
  greater than $2^{n^2}/2$. By Lemma \ref{lem:gatebound},
  worst-case C-circuits require $\Omega(n^2/\log n)$ gates.
}

Let us return to CNT-constructible permutations. A result similar
to Lemma \ref{lem:constr:c} requires:

\thmx{DEF}{ \label{def:alter}
  A permutation is called even if it can be written
  as the product of an even number of transpositions. The set of
  even permutations in $S_n$ is denoted $A_n$.
}

It is well-known that if a permutation can be written as the
product of an even number of transpositions, then it may not be
written as the product of an odd number of transpositions.
Moreover, half the permutations in $S_n$ are even for $n>1$.

\thm{LEM}{ \label{lem:csube} \cite{Toffoli}
  Any $n\times n$ circuit with no $n\times n$ gates
  computes an even permutation.
} {
  It suffices to prove this for a circuit consisting of only one
  gate, as the product of even permutations is even. Let
  $G$ be a gate in an $n \times n$ circuit. By hypothesis, $G$ is
  not $n \times n$, so there must be at least one wire
  which is unaffected by $G$. Without loss of generality,
  let this be the high-order
  wire. Then $2^{n-1}\oplus G(k)=G(2^{n-1}\oplus k)$, and
  $k < 2^{n-1}$ implies $G(k) < 2^{n-1}$. Thus every cycle in the
  cycle decomposition of $G$ appears in duplicate: once with
  numbers less than $2^{n-1}$, and once with the corresponding
  numbers with their high order bits set to one. But these cycles
  have the same length, and so their product is an even
  permutation. Therefore, $G$ is the product of even permutations,
  and hence is even.
}

To illustrate this result, consider the following example. A $2
\times 2$ circuit consisting of a single S gate performs the
permutation $(1,2)$, as the inputs $01$ and $10$ are interchanged,
and the inputs $00$ and $11$ remain fixed. This permutation
consists of one transposition, and is therefore odd. On the other
hand, in a $3 \times 3$ circuit, one can check that a swap gate on
the bottom two wires performs the permutation $(1,2)(5,6)$, which
is even.

\section{Theoretical Results} \label{sec:theory}

Since the CNTS gate library contains no gates of size greater than
three, Lemma \ref{lem:csube} implies that every CNTS-constructible
(without temporary storage) permutation is even for $n\ge 4$. The
main result of this section is that the converse is also true.

\thmx{THM}{ \label{thm:constr:cnt}
  Every even permutation is CNT-constructible.
}

Before beginning the proof, we offer the following two
corollaries. These give a way to synthesize circuits computing odd
permutations using temporary storage, and also extend Theorem
\ref{thm:constr:cnt} to an arbitrary universal gate library.

\thm{COR}{ \label{cor:theoddones}
  Every permutation, even or odd, may be computed in a CNT-circuit
  with at most one wire of temporary storage.
} {
  Suppose we have an $n\times n$ gate G computing $\pi \in S_{2^n}$,
  and we place it on the bottom $n$ wires of an $(n+1)\times(n+1)$
  reversible circuit; let $\tilde{\pi}$ be
  the permutation computed by this new circuit. Then by
  Lemma \ref{lem:csube}, $\tilde{\pi}$
  is even. By Theorem \ref{thm:constr:cnt}, $\tilde{\pi}$ is
  CNT-constructible. Let C be a CNT-circuit computing $\tilde{\pi}$.
  C computes $\pi$ with one line of temporary storage.
}

\thm{COR}{ \label{cor:arbunivgl}
  For any universal gate library $L$ and sufficiently large $n$,
  permutations in $A_{2^n}$ are $L$-constructible, and
  those in $S_{2^n}$ are realizable with at most one
  wire of temporary storage.
} {
  Since $L$ is universal, there is some number $k$
  such that we can compute
  the permutations corresponding to the NOT, CNOT, and TOFFOLI gates
  using a total of $k$ wires. Let $n>k$, and let $\pi \in A_{2^n}$.
  By Theorem \ref{thm:constr:cnt}, we can find a CNT-circuit C computing $\pi$,
  and can replace every N, C, or T gate with a circuit computing it.
  The second claim follows similarly
  from Theorem \ref{thm:constr:cnt} and Corollary \ref{cor:theoddones}.
}

To prove Theorem \ref{thm:constr:cnt}, we begin by asking which
permutations are C-constructible, N-constructible, and
T-constructible. The first of these questions was answered in
Section \ref{sec:background}. We now summarize the properties of
N-constructible permutations. In what follows, $\oplus$ denotes
bitwise XOR.

\thmx{DEF}{ \label{def:ninot}
  Given an integer $i$, we denote by $N^i$ the circuit formed
  by placing an N gate on every wire corresponding to a $1$ in the
  binary expansion of $i$.
}

\begin{figure}[t]
  \begin{center}
    \includegraphics[height=3cm]{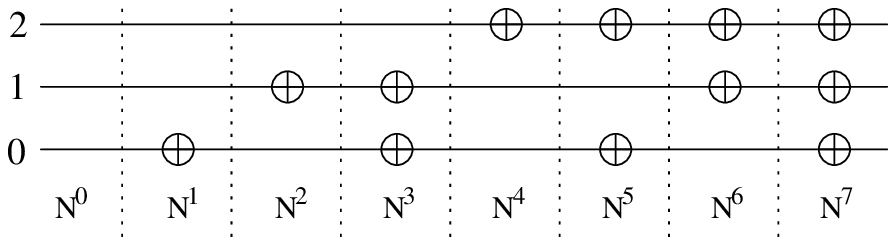}

  \parbox{13cm}
   {
      \caption{
        \label{fig:ninot}
        Circuits $N^i$ for $i < 8$. The superscript is interpreted as a
        binary number, whose non-zero bits correspond to the location of
        inverters.
   }
  }
  \end{center}
\end{figure}

  We will use $N^i$ to signify both the circuit described above,
  and the permutation which this circuit computes. Technically, the latter
  is not uniquely determined by the $N^i$ notation, but also
  depends on the number $n$ of wires in the circuit; however, $n$
  will always be clear from context. The $N^i$ notation is illustrated for
  the case of three wires in Figure \ref{fig:ninot}.

\thm{LEM}{\label{lem:constr:n}
  Let $\pi \in S_{2^n}$ be N-constructible. There exists an
  $i$ such that $\pi(x)=x \oplus i$. Moreover, the gate-minimal circuit
  for $\pi$ is $N^i$. There are $2^n$ N-constructible permutations
  in $S_{2^n}$.
} {
  Clearly, $N^i$ computes the permutation $\pi(x)=x\oplus i$.
  It now suffices to show that an arbitrary N-circuit may be reduced to
  one of the $N^i$ circuits. Any pair of consecutive N gates on
  the same wire
  may be removed without changing the permutation computed by the
  circuit. Applying this transformation until no more gates can be
  removed must leave a circuit with at most one N gate per wire;
  that is, a circuit of the form $N^i$.
}

\subsection{T-Constructible Permutations}

Characterizing the T-constructible permutations is more difficult.
We will begin by extending the $N^i$ notation defined above.

\thmx{DEF}{
  Let $N^h$ be an N-circuit as defined above. Let $k$ be an integer
  such that the bitwise Boolean product $hk=0$. Let there be $p$
  1s in the binary expansion of $h$, and $q$ in the binary expansion
  of $k$. Define $N^h_k$ to be the reversible circuit composed of
  $p$ $q$-CNOT gates, with control bits on the wires specified by
  the binary expansion of $k$, and inverters as specified by the
  binary expansion of $h$. $N^h_k$ performs $N^h$ iff the wires
  specified by $k$ have the value $1$.
}

In a $3 \times 3$ circuit, there are $3$ possible T gates, namely
$N^1_6$, $N^2_5$, and $N^4_3$. They compute the permutations
$(6,7), (5,7), (3,7)$ respectively. By composing these three
transpositions in all possible ways, we may form all $24$
permutations of $3,5,6,7$. These are precisely the non-negative
integers less than $8$ which are not of the form $0$ or $2^i$.
Clearly, no T gate can affect an input with fewer than two $1$s in
its binary expansion.

\thmx{LEM}{ \label{lem:tfix}
  Every T-circuit fixes $0$ and $2^i$ for all $i$.
}

For $k \times k$ T-circuits, $k > 3$, there is an added
restriction. As T gates are $3\times 3$, there can be no $k\times
k$ gates in the circuit, so by Lemma \ref{lem:csube}, the circuit
must compute an even permutation. On the other hand, we will show
that these are the only restrictions on T-constructible
permutations. We will do this by choosing an arbitrary even
permutation, and then giving an explicit construction of a circuit
which computes it using no temporary storage. The first step is to
decompose the permutation into a product of pairs of disjoint
transpositions.

\thm{LEM}{ \label{lem:disjtrans}
  For $n>4$, any even permutation in $S_n$ may be written as the
  product of pairs of disjoint transpositions. If a permutation
  $\pi$ moves $k$ indices, it may be decomposed into no more than
  $\frac{k+1}{2}$ pairs of transpositions.
} {
  By a pair of disjoint transpositions, we mean something of the
  form $(a,b)(c,d)$ where $a,b,c,d$ are distinct. For $k\ge 3$,
  $(x_0, x_1,\ldots, x_k)=(x_0,x_1)(x_{k-1}, x_k)
  (x_0,x_2,x_3,\ldots, x_{k-1})$. Now $(x_0, x_1)(x_{k-1},x_k)$ are
  disjoint, iteratively applying this decomposition process will convert
  an arbitrary cycle into a product of pairs of disjoint
  transpositions possibly followed by a single transposition, a 3-cycle or both.

  Consider an arbitrary permutation $\pi=c_0 c_1 \ldots c_k$,
  where $c_0 \ldots c_k$ are the disjoint cycles in its cycle
  decomposition. As shown above, we may rewrite this as
  $\pi=\kappa_1 \ldots \kappa_m \tau_1 \ldots \tau_p \sigma_1
  \ldots \sigma_q$, where the $\kappa_i$ are pairs of disjoint
  transpositions, the $\tau_i$ are transpositions, and the
  $\sigma_i$ are 3-cycles. As the $\tau_i$ come from pairwise
  disjoint cycles, they must in turn be pairwise disjoint.
  Moreover, there must be an even number of them as $\pi$ was
  assumed to be even, and the $\kappa_i$ and $\sigma_i$ are all even.
  Pairing up the $\tau_i$ arbitrarily leaves an expression of the
  form $\kappa_1 \ldots \kappa_{m+\frac{p}{2}} \sigma_1 \ldots \sigma_q$.
  Again, the $\sigma_i$ are pairwise disjoint. Note that
  $(a,b,c)(d,e,f) = [(a,b)(d,e)][(a,c)(d,f)]$; we may therefore
  rewrite any pair of disjoint 3-cycles as two pairs of disjoint
  transpositions. Iterating this process leaves at most one
  3-cycle, $(x,y,z)$. Since we are working in $A_n$ for $n>4$,
  there are at least two other indices, $v,w$. Using these, we
  have $(x,y,z)=[(x,y)(v,w)][(v,w)(x,z)]$.

  A careful count of transposition pairs gives the bound $\frac{k+1}{2}$ in
  the statement of the lemma. This bound is tight in the case of a permutation
  consisting of a single $4n+3$ cycle.
}

By Lemma \ref{lem:disjtrans}, it suffices to show that we may
construct a circuit for an arbitrary disjoint transposition pair.
We begin with an important special case. On $n$ wires, a
$N^1_{2^k-4}$ gate computes the permutation $\kappa_0 = (2^n
-4,2^n -3)(2^n -2,2^n -1)$, which may be implemented by $8(n-5)$ T
gates \cite[Corollary 7.4]{Barenco}.

\thmx{LEM}{\label{lem:specdisjtrans}
  On $n$ wires, the permutation $\kappa_0 = (2^n -4,2^n -3)(2^n -2,2^n -1)$
  is T-constructible.
}

Consider now an arbitrary disjoint transposition pair, $\kappa =
(a,b)(c,d)$. Given a permutation $\pi$ with the property
$\pi(a)=2^n-4$, $\pi(b)=2^n-3$, $\pi(c)=2^n-2$, $\pi(d)=2^n-1$, we
have $\kappa = \pi \kappa_0 \pi^{-1}$, where $\kappa_0$ is the
permutation in Lemma \ref{lem:specdisjtrans}. We have a circuit
which computes $\kappa_0$. Given a circuit that computes $\pi$, we
may obtain a circuit computing $\pi^{-1}$ by reversing it. We now
construct a circuit computing $\pi$.

\thm{LEM}{ \label{lem:specdisjtransconj}
  Suppose $n>3$, and $0 \le a,b,c,d < 2^n$. Further suppose that
  none of $a,b,c,d$ is $0$, or of the form $2^i$. Then there
  exists a T-constructible permutation $\pi$ with the property
  $\pi(a)=2^n-1$, $\pi(b)=2^n-2$, $\pi(c)=2^n-3$, $\pi(d)=2^n-4$,
  computable by a circuit of no more than $5n-2$ T gates.
} {
  To simplify notation, set $M=2^{n-1}$ and $m=n-1$.
  Now, we construct $\pi$ in five stages. First, we build a
  permutation $\pi_a$ such that $\pi_a(a)=M+4$. Then, we build
  $\pi_b$ such that $\pi_b \circ \pi_a(b)=M+1$, and
  $\pi_b(M+4)=M+4$. Similarly, $\pi_c$ will fix $M+1$ and
  $M+4$, while $\pi_c \circ \pi_b \circ \pi_a (c) = M+2$, and
  $\pi_d$ will fix $M+1$, $M+2$, $M+4$ while
  $\pi_d \circ \pi_c \circ \pi_b \circ \pi_a (d) = M+7$.
  Finally, we build a circuit that maps $M+4 \mapsto 2M - 4$,
  $M+1 \mapsto 2M-3$, $M+2 \mapsto 2M-2$, and $M+7
  \mapsto 2M -1$.

  By hypothesis, $a$ is not $0$ or of the form $2^i$. This means
  that $a$ has at least two $1$s in its binary expansion, say in
  positions $h_a$ and $k_a$. Apply T gates with controls on positions
  $h_a$ and $k_a$ to set the second and $m$th bits. More precisely,
  let $z_a = 2^{h_a} + 2^{k_a}$, apply a $N^M_{z^a}$
  iff $a$ has a $0$ in the $(n-1)$st bit
  and $N^4_{z_a}$ iff $a$ has a $0$ in the $2$nd bit. Now,
  apply T gates with the controls on the $m$th and $2$nd bits
  to set the remaining bits to $0$.
  Let $\pi_a$ be the permutation computed by the
  circuit given above.

  $\pi_a(b)$ must again have two nonzero bits in its binary
  expansion; since $b \ne a$ implies $\pi_a(b) \ne \pi_a(a)$, some
  nonzero bit of $\pi_a(b)$ lies on neither the $m$th nor the
  $2$nd wire. Controlling by this and another bit, use the
  techniques of the previous paragraph to build a circuit taking
  $\pi_a(b) \to M + 1$. By construction, this fixes $M + 4$;
  let the permutation computed by this circuit be $\pi_b$.

  Consider now the nonzero bits of $c'=\pi_b \circ \pi_a(c)$. Again,
  since $a,b \ne c$, we have $M+4,M+1 \ne c'$.
  Therefore, there must be at least one bit in which
  $c'$ differs from $M + 4$. This bit could be the $m$th or the second bit,
  and $c'$ could have a zero in this position.
  However, as $c'$ is guaranteed to have at least $2$
  non-zero bits, there must be some other bit which is $1$ in $c'$
  and $0$ in $M + 4$. Similarly, there must be some bit
  which is $1$ in $c'$ and $0$ in $M + 1$. Controlling by
  these two bits (or, if they are the same bit, by this bit and any
  other bit which is $1$ in $c'$), we may use the above method to
  set $c' \to M+2$.

  Next, consider the nonzero bits of $d'=\pi_c \circ \pi_b \circ
  \pi_a(d)$. First, suppose there are two which are not
  on the $m$th wire. Controlling by these can take $d' \to
  M+ 7$ without affecting any of the other values, as none of
  $M+1,M+2, M+4$ have $1$s in both these positions.
  If there are no two $1$s in the binary expansion of $d'$ which
  both lie off the $m$th wire, there can be at most two $1$s in
  the binary expansion, one of which lies on the $m$th wire. Since
  $a,b,c \ne d$, the second must lie on some wire which is not the
  $0$th, $1$st, or $2$nd; in this case we may again control by
  these two bits to take $d' \to M + 7$ without affecting
  other values.

  Finally, apply $N^4_{M+1}$ and $N^4_{M+2}$
  gates, and then a $N^{M-8}_{M+4}$ circuit.
  The reader may verify that this completes stage 5. Each of the
  first 4 stages takes at most $n$ T gates, as we flip at most $n$
  bits in each. The final stage uses exactly $n-2$ T gates.
}

We now have a key result to prove.

\thm{THM}{\label{thm:constr:t}
  Every T-constructible permutation in $S_{2^n}$ fixes $0$ and
  $2^i$ for all $i$, and is even if $n>3$. Conversely,
  every permutation of this form is T-constructible. A T-constructible
  permutation which moves $s$ indices requires at most
  $3(s+1)(3n-7)$ T gates. There
  are $\frac{1}{2} (2^{n}-n-1)!$ T-constructible permutations in
  $S_{2^n}$.
}{
  We have already dealt with the case $n=3$; hence suppose $n>3$.
  The first statement follows directly from Lemmas \ref{lem:csube}
  and \ref{lem:tfix}. Now let $\pi \in S_{2^n}$ be an
  arbitrary even permutation fixing $0$, $2^i$. Use the method
  of Lemma \ref{lem:disjtrans} to decompose $\pi$ into pairs of
  disjoint transpositions which fix $0$, $2^i$. We are justified
  in using Lemma \ref{lem:disjtrans}
  because, for $n > 3$, there are at least five numbers between
  $0$ and $2^{n-1}$ which are not of the form $0$ or $2^i$.
  Finally, using the circuits implied by Lemmas \ref{lem:specdisjtrans} and
  \ref{lem:specdisjtransconj}, we may construct circuits
  for each of these transposition pairs. Chaining these circuits together
  gives a circuit for the permutation $\pi$. Collecting the
  length bounds of the various lemmas cited gives the length bound
  in the theorem. The final claim then follows.
}

\subsection{Circuit Equivalences}

Given a (possibly long) reversible circuit to perform a specified
task, one approach to reducing the circuit size is to perform
local optimizations using circuit equivalences. The idea is to
find subcircuits amenable to reduction. This direction is pursued
in a paper by Iwama et al. \cite{Japanese}, which examines
circuit transformation rules for generalized-CNOT circuits which
only alter one bit of the circuit. In their scenario, other bits
may be altered during computation, so long as they are
returned to their initial state by the end of the computation. We
present a more general framework for deriving equivalences, from which
many of the equivalences from \cite{Japanese} follow as special cases.
First, let us introduce notation to better deal with control bits.

\thmx{DEF}{\label{def:control}
  Let $G^i$ be a reversible gate that only affects wires
  corresponding to the $1$s in the binary expansion of $i$
  (as in an $N^i$ gate). Let the bitwise Boolean product $i \cdot j
  = 0$. Then define $V_j (G^i)$ as the gate which computes $G^i$
  iff the wires specified by $j$ all carry a $1$.
}

In particular, $V_j(N^i) = N^i_j$, and $V_k V_j (G^i) = V_{k+j}
(G^i)$. Addition, multiplication, etc., of lower indices will
always be taken to be bitwise Boolean, with $+$, $\cdot$, $\oplus$
representing OR, AND, and XOR respectively. We denote the bitwise
complement of $x$ as $\overline{x}$.

\thm{LEM}{\label{lem:contlift}
  Let $K$ be an $n \times n$ reversible circuit such that $K(0 x_1 \ldots
  x_{n-1}) = (0 x_1 \ldots x_{n-1})$, and let $f:B^{n-1} \to B^{n-1}$ be the
  function defined by $K(1 x_1 \ldots x_{n-1}) = (1 f(x_1 \ldots
  x_{n-1}))$. Then $f$ is a well-defined permutation in
  $S_{2^{n-1}}$, and if $F$ is a circuit computing $f$, then
  $V_1 (F) \equiv K$.
} {
  $K$, by hypothesis, permutes the inputs with a leading $0$
  amongst themselves. By reversibility, it must permute inputs with a
  leading $1$ amongst themselves as well.
}

\thmx{DEF}{\label{def:commutator} The {\em commutator} of
permutations $P$ and $Q$, denoted $[P,Q]$, is $PQP^{-1}Q^{-1}$.}

\vspace{-2mm}
The commutator concept is useful for moving gates past each other
since $PQ = [P,Q]QP$. Moreover, it has reasonable properties with respect to
control bits as the following result indicates.

\thm{COR}{\label{cor:commcontlift}
  $[V_h(G^i), V_k(H^j)]=V_{(h+k)\cdot\overline{(i+j)}}([V_{h\cdot j}(G^i),
  V_{k\cdot i}(H^j)])$
} { The corollary provides a circuit equivalent to the commutator
    of two given gates with arbitrary control bits.
    Namely, such a circuit can be constructed in two steps.
    First, identify wires which act as control for one gate
    but are not touched by the other gate.
    Second, connect the latter gate to every such wire so that
    the wire controls the gate.

By induction, it suffices to show that this procedure can be done to
one such wire. Without loss of generality, suppose control bits
and only control bits appear on the first wire. Then the input to
this wire goes through the circuit unchanged. At least one of the
two gates whose commutator is being computed must, by hypothesis,
be controlled by the first wire. Therefore, on an input of zero to
the first wire, this gate (and therefore its inverse) leaves all
signals unchanged. Since the other gate appears along with its
inverse, the whole circuit leaves the input unchanged. Our result
now follows from Lemma \ref{lem:contlift}.
}

\vspace{-2mm} If we are computing the commutator of generalized
CNOT gates, then we may pick $G^i, H^j$ to be single inverters
$N^i, N^j$ with $i,j$ having only a single $1$ apiece in their
binary expansions. Then we must have $h\cdot j=0$ or $j$, and
$k\cdot i=0$ or $i$. The four cases are accounted for as follows:

\vspace{-2mm} \thm{LEM}{\label{lem:commnot}
  Let $i,j$ have only a single $1$ apiece in their binary expansions. Then
  $[N^i,N_i^j]=N^j$, $[N^i_j,N^j]=N^i$, $[N^i, N^j]= 1$, and
  $[N^i_j,N^j_i]=N^j_i$.
} {
  As these equivalences all involve only 2-bit circuits, we may
  check them for $i=0$, $j=1$ by evaluating
  both sides of each equivalence on each of $4$ inputs.
}

\subsection{CT$|$N and C$|$T Constructible Permutations}

While an arbitrary CNT-circuit may have the C, N, and T gates
interspersed arbitrarily, we first consider circuits in which
these gates are segregated by type.

\thmx{DEF}{ \label{def:constr2}
  For any gate libraries $L_1 \ldots L_k$,
  a $L_1 | \ldots | L_k$-circuit is an
  $L_1$-circuit followed by an $L_2$-circuit, \ldots, followed
  by an $L_k$-circuit. A permutation computed by an $L_1|\ldots|L_k$-circuit
  is $L_1| \ldots | L_k$-constructible.
}

A CNT-circuit with all N gates appearing at the right end is called
a CT$|$N circuit.

\begin{figure}\begin{center}
    \includegraphics[height=2.3cm]{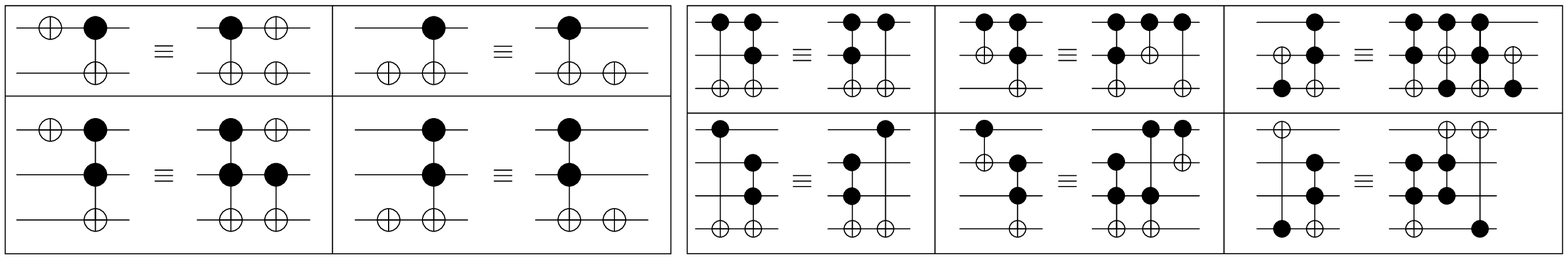}\\
    \hspace{-16mm} (a) \hspace{62.7mm} (b)
  \vspace{-2mm}
  \caption{ \label{fig:movecnt}
    Equivalences between reversible circuits
    used in our constructions.
  }
  \end{center}
\end{figure}

\thm{THM}{ \label{thm:constr:ct:n}
  Let $\pi$ be CNT-constructible. Then $\pi$ is also
  CT$|$N-constructible. Moreover, $\pi$ uniquely determines the
  permutations $\pi_{CT}$ and $\pi_N$ computed by the CT and N
  sub-circuits, respectively.
} {
  We move all the N gates toward the outputs of the circuit.
  Each box in Figure \ref{fig:movecnt}$a$ indicates a way of replacing
  an N$|$CT circuit with a CT$|$N circuit. The equivalences in
  this figure come from Corollary \ref{cor:commcontlift}.
  Moreover, every possible way for an N gate
  to appear to the immediate left of a C or a T is accounted for,
  up to permuting the input and output wires.
  Now, number the non-N gates in the circuit in a reverse topological order
  starting from the outputs. In particular, if two gates appear at the same
  level in a circuit diagram,
  they must be independent, and one can order them arbitrarily.
  Let $d$ be the number of the highest-numbered gate with an N gate
  to its immediate left.
  All N gates past the $d$-th gate $G$ can be reordered with the $G$ gate
  without introducing new N gates on the other side of $G$,
  and without introducing new gates between the N gates and the outputs.
  In any event, as there are no remaining N gates to the left of $G$,
  $d$ decreases. This process terminates with all the N gates are clustered
  together at the circuit outputs.
  If we always cancel redundant pairs of N gates, then no more than two
  new gates will be introduced for each non-inverter originally in the circuit;
  additionally, there will be at most $n$ N gates when the process
  is complete. Thus if the original circuit had $l$ gates,
  then the new circuit
  has at most $3(l-1)+n$ gates.
  Note that C and T gates (and hence CT-circuits) fix 0. Thus
  $\pi(0)=\pi_N(0)$, so $\pi_N = N^{\pi(0)}$, and
  $\pi_{CT}=\pi N^{\pi(0)}$.
}

Thus, if we want a CNT-circuit computing a permutation $\pi$, we
can quickly compute $\pi_N$ and then simplify the problem to that
of finding a CT-circuit for $\pi \pi_N$. By Theorem
\ref{thm:constr:ct:n}, we know that a minimal-gate circuit of this
form has roughly three times as many gates as the gate-minimal
circuit computing $\pi$.

The next natural question is whether an arbitrary CT-circuit is
equivalent to some T$|$C circuit. The equivalences in Figure
\ref{fig:movecnt}$b$ suggest that the answer is yes. However, the
proof of Theorem \ref{thm:constr:ct:n} requires that many N gates
be able to simultaneously move past a C or T gate, while Figure
\ref{fig:movecnt} only shows how to move a single C gate past a
single T gate.

\thm{LEM}{\label{lem:constr:t:c}
  The permutation $\pi$ computed by a T$|$C-circuit determines the
  permutations $\pi_T$ and $\pi_C$ computed by the sub-circuits.
  An even permutation is TC-constructible iff it fixes $0$
  and the images of inputs of the form $2^i$ are linearly independent
  over $\mathbb{F}_2$.
} {
  Let $\pi$ be an arbitrary permutation. If $\pi$ is
  T$|$C-constructible, then images of the inputs $2^i$ are
  unaffected by the T subcircuit; by Lemma \ref{lem:constr:c} they must
  be mapped to linearly independent values by the C subcircuit.
  This mapping of basis vectors completely specifies the
  permutation $\pi_c$ computed by the C subcircuit, and therefore
  also the permutation $\pi_t = \pi \pi_c^{-1}$ computed by the
  T subcircuit. Conversely, suppose $\pi$ is even and fixes $0$,
  and the images of $2^i$ are linearly independent. Then there
  is some C-circuit taking the values $2^i$ to their images
  under $\pi$. Let it compute the permutation $\pi_c$; then
  $\pi \pi_c^{-1}$ fixes the values $0$ and $2^i$ by construction.
  Theorem \ref{thm:constr:t} therefore guarantees that
  $\pi \pi_c^{-1}$ is T-constructible.
}

We will later use this result to show the existence of
CT-constructible permutations which are not T$|$C constructible.

\subsection{T$|$C$|$T$|$N-Constructible Permutations}

We are now ready to prove Theorem \ref{thm:constr:cnt}.
According to Lemma \ref{lem:constr:t:c}, zero-fixing even
permutations are T$|$C-constructible if they map inputs of the
form $2^i$ in a certain way. This suggests that T$|$C-circuits
account for a relatively large fraction of such permutations.

\vspace{-2mm} \thm{THM}{ \label{thm:constr:t:c:t}
  Every zero-fixing permutation in $S_{2^3}$ and every
  zero-fixing even permutation in
  $S_{2^n}$ for $n>4$ is T$|$C$|$T-constructible, and hence is
  CT-constructible. None requires more than $n^2$ C gates
  and $3(2^n +n +1)(3n-7)$ T gates.
} {
  Let $\pi$ be any zero-fixing permutation.
  Note that if the images of $2^i$ under $\pi$ were linearly
  independent, Lemma \ref{lem:constr:t:c} would imply that $\pi$ was
  T$|$C constructible. So, we will build a permutation $\pi_T$
  with the property that the images of $2^i$ under $\pi \pi_T$
  are linearly independent, ensuring that $\pi \pi_T$ is
  T$|$C-constructible. Given a T$|$C-circuit for $\pi \pi_T$ and a
  T-circuit for $\pi_T$, we can reverse the circuit for $\pi_T$
  and append it to the end of the T$|$C-circuit for $\pi \pi_T$ to
  give at T$|$C$|$T-circuit for $\pi$. All that remains is to show
  we can build one such $\pi_T$.

  The basis vectors $2^i$ must be mapped either to themselves, to
  other basis vectors, or to vectors with at least two $1$s. Let
  $i_1 \ldots i_k$ be the indices of basis vectors which are not
  the images of other basis vectors, and let $j_1 \ldots j_k$ be
  the indices of basis vectors whose images have at least two
  $1$s. Let $\bar{i}_1 \ldots \bar{i}_{n-k}$ and $\bar{j}_1 \ldots
  \bar{j}_{n-k}$ be the indices which are not in
  the $i_m$ and $j_m$ respectively. Consider the matrix $M_\pi$
  in which the $i$th column is the binary expansion of $\pi(2^i)$.
  We take the entries of $M_\pi$ to be elements of $\mathbb{F}_2$.
  Our indexing system divides $M_\pi$ into four submatrices; $M_\pi(i,j)$,
  $M_\pi(i,\bar{j})$, $M_\pi(\bar{i},j)$, and $M_\pi(\bar{i},\bar{j})$. By
  construction, $M_\pi(i,j)$ and $M_\pi(\bar{i},\bar{j})$ are square,
  $M_\pi(\bar{i},\bar{j})$ is a permutation matrix, and $M_\pi(i,\bar{j})$
  is a zero matrix. Therefore, $\det{M_\pi} = \det{M_\pi(i,j)}$, and
  $M_\pi$ is invertible iff $M_\pi(i,j)$ is. Moreover, there is an
  invertible linear transformation, computable by column-reduction,
  which zeroes out the matrix $M_\pi(\bar{i},j)$
  without affecting $M_\pi(i,j)$ or $M_\pi(\bar{i},\bar{j})$. As this
  transformation $L$ is invertible, it corresponds to a permutation
  $\pi_x$, and the matrix $ML$ is the matrix of images of $2^i$
  under the permutation $\pi_x \pi$. In particular, the columns of
  $(ML)_\pi$ must all be different, which implies that the columns of
  $M_\pi(i,j)$ must all be different. Moreover,
  $\pi_x$ is linear, and therefore zero-fixing; hence $M_\pi(i,j)$ can
  have no zero columns. Taken together, these facts imply that for
  $k=1,2$, $M_\pi(i,j)$ is invertible, hence so is $M_\pi$, thus
  $\pi$ is T$|$C-constructible.

  Suppose $k\ge 3$, and consider the family of matrices $A(p)$
  defined as follows. $A(p)$ is a $p \times p$ matrix with $1$s on
  the diagonal, $1$s in the first row, and $1$s in the first
  column, except possibly in the $(1,1)$ entry, which is $1$ iff
  $p$ is odd. Row-reducing the $A_i$ to lower triangular matrices
  quickly shows that the $A_i$ are invertible for all $i$.
  Moreover, for $i\ge 3$, there is at least two $1$s in every
  column. Therefore, there is a T-constructible permutation
  $\pi_T$ such that $M_{\pi \pi_T}(i,j)=A_k$. Thus $\pi \pi_T$ is
  T$|$C-constructible, and $\pi$ is T$|$C$|$T constructible.

  Finally, we know from Corollary \ref{cor:cnotcount} that no more
  than $n^2$ gates are necessary to compute $\pi_C$.
  At most $2n$ indices need be moved by $\pi_T$, and no more than
  $2^n - n - 1$ can be moved by the T-constructible part of $\pi$.
  Thus by Theorem \ref{thm:constr:t}, we need no more than
  $3(2n+1)(3n-7)$ gates for $\pi_T$ and no more than $3(2^n -
  n)(3n-7)$ gates for $\pi$. Adding these gives the gate-count estimate
  above.
}

\addtolength{\baselineskip}{1mm} \thm{COR}{
\label{cor:constrcx:t:c}
  There exist T$|$C$|$T-constructible permutations which are not
  T$|$C-constructible.
}{
  The permutation $\pi=(2,6)(4,7)$ fixes $0$ and is even, hence
  is T$|$C$|$T-constructible in $S_{2^n}$ for all $n\ge 3$ by
  Theorem \ref{thm:constr:t:c:t}. However,
  $\pi(1)\oplus \pi(2) = 1 \oplus 6 = 7 = \pi(4)$, hence by Lemma
  \ref{lem:constr:t:c}, $\pi$ is not T$|$C-constructible.
}

\thm{THM}{ \label{thm:constr:t:c:t:n}
  Every permutation in $S_{2^n}$ for $n = 1,2,3$ and every even
  permutation in $S_{2^n}$ for $n>3$ is T$|$C$|$T$|$N-constructible,
  and hence CNT-constructible. None requires more than $n^2$ C
  gates, $n$ N gates, and $3(2^n +n +1)(3n-7)$ T gates.
} {
  Let $\pi$ be any permutation; then $\pi' = \pi N^{\pi(0)}$ fixes $0$.
  For $n=1$, $\pi'$ must be the identity; for $n=2$ $\pi'$ permutes
  $1,2,3$, any such permutation is linear, hence $\pi'$ is
  C-constructible. For $n=3$, $\pi'$ is T$|$C$|$T-constructible;
  for $n>3$, $\pi'$ is T$|$C$|$T-constructible iff it is even,
  which happens iff $\pi$ is even. Thus in all cases there is a
  T$|$C$|$T-circuit, $\Pi'$ computing $\pi'$; then $\Pi' N^{\pi(0)}$
  is a T$|$C$|$T$|$N-circuit computing $\pi$.
}

We note that the size of a truth table for a circuit with $n$
inputs and $n$ outputs is $n2^n$ bits. The synthesis procedure
used in the theorems above clearly runs in time proportional to
the number of gates in the final circuit. This is $O(n 2^n)$,
hence the synthesis procedure detailed in the theorems has linear
runtime in the input size.

Just as in Corollary \ref{cor:cnotcount}, we may ask how far from
optimal the foregoing construction is for long circuits. There are
$2^n ! / 2$ even permutations in $S_{2^n}$, and these are all
CNT-constructible. Using Stirling's approximation, $\log(k!)
\approx k \log k$, and Lemma \ref{lem:gatebound} gives:

\thmx{COR}{ \label{cor:t:c:t:n:count}
  Worst case CNT-circuits on $n$ wires require $\Omega(n 2^n/\log n)$
  gates.
}

So, for long CNT-circuits, the algorithm implied by Theorem
\ref{thm:constr:t:c:t:n} is asymptotically suboptimal by, at
worst, a logarithmic factor, as it produces circuits of length
$O(n 2^n)$. This is remarkably similar to the result of Corollary
\ref{cor:cnotcount}, in which we found that using row reduction to
build C-circuits is asymptotically suboptimal by a logarithmic
factor in the case of long C-circuits. However, even a constant
improvement in size is very desirable, and circuits for practical
applications are almost never of the worst-case type considered in
Corollaries \ref{cor:cnotcount} and \ref{cor:t:c:t:n:count}.

\addtolength{\baselineskip}{-1mm}
\newpage
\section{Optimal Synthesis} \label{sec:algos}

We will now switch focus, and seek {\em optimal} realizations for
permutations we know to be CNT-constructible. A circuit is optimal
if no equivalent circuit has smaller cost; in our case, the cost
function will be the number of gates in the circuit.

\thm{LEM}{
  (Property of Optimality) If $B$ is a
  sub-circuit of an optimal
  circuit $A$, then $B$ is optimal. \label{lem:opt}
} {
  Suppose not. Then let $B'$ be a circuit with fewer
  gates than $B$, but computing the same function.
  If we replace $B$ by $B'$, we get
  another circuit $A'$ which computes the same function as $A$. But since we
  have only modified $B$, $A'$ must be as much smaller than $A$
  as $B'$ is smaller than $B$. $A$ was assumed to be optimal, hence
  this is a contradiction.
  (Note that equivalent, optimal circuits can have the same number of gates.)
}

The algorithm detailed in this section relies entirely on the
property of optimality for its correctness. Therefore, any
cost function for which this property holds may, in principle, be
used instead of gate count.

Lemma \ref{lem:opt} allows us to build a library of small optimal
circuits by dynamic programming because the first $m$ gates of an
optimal $(m+1)$-gate circuit form an optimal subcircuit.
Therefore, to exa\-mine all optimal $(m+1)$-gate circuits, we
iterate through optimal $m$-gate circuits and add single gates at
the end in all possible ways. We then check the resulting circuits
against the library, and eliminate any which are equivalent to a
smaller circuit. In fact, instead of storing a library of all
optimal circuits, we store one optimal circuit per synthesized
permutation and also store optimal circuits of a given size
together.

\begin{figure}[!th]
  \begin{center}
    \begin{tabular}{|l|}
      \hline
      \\
      {\tt CIRCUIT find\_circ(COST, PERM)}         \\
      // assumes circuit library stored in LIB \\
      \\
      {\tt if (COST $\le$ k) } \\
      \\
      ~~    // If PERM can be computed by a circuit with $\leq$ k gates, \\
      ~~    // such a circuit must be in the library \\
      {\tt~~    return LIB[DEPTH].find(PERM)}\\
      \\
      {\tt else }   \\
      \\
      ~~    // Try building the goal circuit from $\leq$k-gate circuits \\
      {\tt~~    for each C in LIB[k]}\\
      \\
      ~~~~      // Divide PERM by permutation computed by C \\
      {\tt~~~~      PERM2 $\leftarrow$ PERM * INVERSE(C.perm)}\\
      \\
      ~~~~      // and try to synthesize the result \\
      {\tt~~~~      TEMP\_CCT $\leftarrow$ find\_circ(depth-k,PERM2)}\\
      {\tt~~~~      if (TEMP\_CCT != NIL) return TEMP\_CCT * C}\\
      \\
      // Finally, if no circuit of the desired depth can be found\\
      {\tt return NIL}\\
      \\
      \hline
    \end{tabular}
 \parbox{13.8cm}
 {
  \caption{\label{fig:findcct}
    Finding a circuit of cost $\leq$COST that computes permutation PERM
    (NIL returned if no such circuit exists). TEMP\_CCT and records in LIB
    represent circuits, and include a field ``perm'' storing
    the permutation computed. The * character means both multiplication of
    permutations and concatenation of circuits, and NIL*$<$anything$>$=NIL.
    }
    \vspace{-8mm}
 }
 \end{center}
\end{figure}

One way to find an optimal circuit for a given permutation $\pi$
is to generate all optimal $k$-gate circuits for increasing values
of $k$ until a circuit computing $\pi$ is found. This procedure
requires $\Theta(2^n!)$ memory in the worst case ($n$ is the
number of wires) and may require more memory than is available.
Therefore, we stop growing the circuit library at $m$-gate
circuits, when hardware limitations become an issue. The second
stage of the algorithm uses the computed library of optimal
circuits and, in our implementation, starts by reading the library
from a file. Since little additional memory is available, we trade
off runtime for memory.

We use a technique known as {\em depth-first search with iterative
deepening} (DFID) \cite{Korf}. After a given permutation is
checked against the circuit library, we seek circuits with $j=m+1$
gates that implement this permutation. If none are found, we seek
circuits with $j=m+2$ gates, etc. This algorithm, in general,
needs an additional termination condition to prevent infinite
looping for inputs which cannot be synthesized with a given gate
library. For each $j$, we consider all permutations optimally
synthesizable in $m$ gates. For each such permutation $\rho$, we
multiply $\pi$ by $\rho^{-1}$ and recursively try to synthesize
the result using $j-m$ gates. When $j-m\leq m$, this can be done
by checking against the existing library. Otherwise, the recursion
depth increases. Pseudocode for this stage of our algorithm is
given in Figure \ref{fig:findcct}.

\begin{table}[!th]
  \small
  \begin{center}
    \begin{tabular}{|r|r|r|r|r|r|r|r|r|}\hline
      Size & N & C  & T & NC  & CT   & NT    & CNT & CNTS\\
      \hline
      12    & 0 & 0   & 0  & 0    & 0    & 47    & 0   & 0\\
      11    & 0 & 0   & 0  & 0    & 0    & 1690  & 0   & 0\\
      10    & 0 & 0   & 0  & 0    & 0    & 8363  & 0   & 0\\
      9     & 0 & 0   & 0  & 0    & 0    & 12237 & 0   & 0\\
      8     & 0 & 0   & 0  & 0    & 6    & 9339  & 577 & 32 \\
      7     & 0 & 0   & 0  & 14   & 386  & 5097  & 10253 & 6817\\
      6     & 0 & 2   & 0  & 215  & 1688 & 2262  & 17049 & 17531\\
      5     & 0 & 24  & 0  & 474  & 1784 & 870   & 8921 & 11194\\
      4     & 0 & 60  & 5  & 393  & 845  & 296   & 2780 & 3752 \\
      3     & 1 & 51  & 9  & 187  & 261  & 88    & 625 & 844 \\
      2     & 3 & 24  & 6  & 51   & 60   & 24    & 102 & 134 \\
      1     & 3 & 6   & 3  & 9    & 9    & 6     & 12 & 15 \\
      0     & 1 & 1   & 1  & 1    & 1    & 1     & 1 & 1\\ \hline
      Total & 8 & 168 & 24 & 1344 & 5040 & 40320 & 40320 & 40320\\
      \hline
      Time  &  1&   1 &  1 & 30   & 215  & 97    & 40    & 15\\
      \hline
    \end{tabular}
  \parbox{15cm}
  {
      \caption
           {
        Number of permutations computable in an optimal
        $L$-circuit using a given number of gates. $L \subset CNTS$.
        Runtimes are in seconds for a 2GHz Pentium-4 Xeon CPU.
        \label{tab:3wcctl}
      }
     \vspace{-9mm}
   }
  \end{center}
\end{table}

 In addition to being more memory-efficient than straightforward
dynamic programming, our algorithm is faster than branching over
all possible circuits. To quantify these improvements, consider a
library of circuits of size $m$ or less, containing $l_m$ circuits
of size $m$. We analyze the efficiency of the algorithms discussed
by simulating them on an input permutation of cost $k$. Our
algorithm requires $l_m^{\lfloor(k-1)/m \rfloor}$ references to
the circuit library. Simple branching is no better than our
algorithm with $m=1$, and thus takes at least $l_1^k$ steps, which
is $l_1^k/l_m^{\lfloor(k-1)/m \rfloor}$ times more than our
algorithm. A speed-up can be expected because $l_m\leq l_1^m$, but
specific numerical values of that expression depend on the numbers
of suboptimal and redundant optimal circuits of length $m$.
Indeed, Table  \ref{tab:3wcctl} lists values of $l_m$ for various
subsets of the CNTS gate library and $m=3$. For example, for the
NT gate library, $k=12$, $\lfloor(k-1)/m \rfloor=3$, $l_1=6$ and
$l_m=88$. Therefore the performance ratio is
$l_1^k/l_m^{\lfloor(k-1)/m \rfloor}=6^{12}/88^3\approx 3194.2$.
Yet, this comparison is incomplete because it does not account for
time spent building circuit libraries. We point out that this
charge is amortized over multiple synthesis operations. In our
experiments, generating a circuit library on three wires of up to
three gates ($m=3$) from the CNTS gate library takes less than a
minute on a 2-GHz Pentium-4 Xeon. Using such libraries, all of
Table \ref{tab:3wcctl} can be generated in minutes,\footnote{
Although complete statistics for all 16! 4-wire functions are
beyond our reach, average synthesis times are less than one second
when the input function can be implemented with eight gates or
fewer. Functions requiring nine or more gates tend to take more
than 1.5 hours to synthesize. In this case, memory constraints
limit our circuit library to 4-gate circuits, and the large jump
in runtime after the 8-gate mark is due to an extra level of
recursion. } but it cannot be generated even in several hours
using branching.

\begin{figure}[!t]
\begin{center}
    \includegraphics[width=14cm]{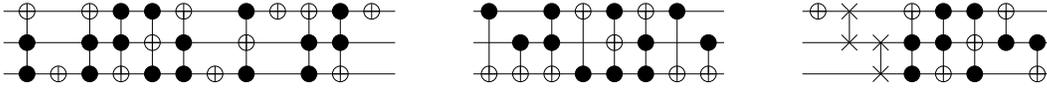}
  \end{center}
  \caption{
    Worst-case $L$-circuits where $L$ is NT, CNT and CNTS.
    \label{fig:ntmaxgates}
  }
\end{figure}

Let us now see what additional information we can glean from Table
\ref{tab:3wcctl}.  Adding the C gate to the NT library appears to
significantly reduce circuit size, but further adding the S gate
does not help as much. To illustrate this, we show sample
worst-case circuits on three wires for the NT, CNT, and CNTS gate
libraries in Figure \ref{fig:ntmaxgates}.

The totals in Table \ref{tab:3wcctl} can be independently
determined by the following arguments.  Every reversible function
on three wires can be synthesized using the CNT gate library
\cite{Toffoli}, and there are $8!=40,320$ of these. All can be
synthesized with the NT library because the C gate is redundant in
the CNT library; see Figure \ref{fig:moreeq}$a$. On the other
hand, adding the S gate to the library cannot decrease the number
of synthesi\-zable functions. Therefore, the totals in the NT and
CNTS columns must be $40,320$ as well. On the other side of the
table, the number of possible N circuits is just $2^3=8$ since
there are three wires, and there can be at most one N gate per
wire in an optimal circuit (else we can cancel redundant pairs.)
By Theorem \ref{thm:constr:ct:n}, the number of CN-constructible
permutations should be the product of the number of
N-constructible permutations and the number of C constructible
permutations, since any CN-constructible permutation can be
written uniquely as a product of an N-constructible and a
C-constructible permutation. So the total in the CN column should
be the product of the totals in the C and N columns, which it is.
Similarly, the total in the CNT column should be the product of
the totals in the CT and N columns; this allows one to deduce the
total number of CT-constructible permutations from values we know.
Finally, we showed that there were $24$ T-constructible
permutations on 3 wires in Section \ref{sec:theory}, and Corollary
\ref{cor:cnotcount} states that the number of permutations
implementable on $n$ wires with C gates is $\prod_{i=0}^{n-1}
(2^n-2^i)$. For $n=3$ this yields 168 and agrees with Table
\ref{tab:3wcctl}.

We can also add to the discussion of T$|$C constructible circuits
we began in Section \ref{sec:theory}. By Lemma
\ref{lem:constr:t:c}, the number of T$|$C-constructible
permutations can be computed as the product of the numbers of
T-constructible and C-constructible permutations. Table
\ref{tab:3wcctl} mentions $24$ T-circuits and $168$ C-circuits on
three wires. The product, $4032$, is less than $5040$, the number
of CT constructible permutations on three wires, as we would
expect from Corollary \ref{cor:constrcx:t:c}.

Finally, the longest C-circuits we observed on 3, 4 and 5 wires
merely permute the wires. Such wire-permutations on $n$ wires
never require more than $3(n-1)$ gates. However, from Corollary
\ref{cor:cnotcount} we know that for large $n$, worst-case
C-circuits require $\Omega(n^2/log(n))$ gates. Identifying
specific worst-case circuits and describing families with
worst-case asymptotics remains a challenge.

Finally, we note that while the exact runtime complexity of this
algorithm is dependant on characteristics of the gate library
chosen, for a complete gate library it is obviously exponential in
the number of input wires to the circuit (this is guaranteed by
Corollary \ref{cor:t:c:t:n:count}), and in fact must be at least
doubly-exponential in the number of input wires (that is,
exponential in the size of the truth table). Scalability issues,
therefore, restrict this approach to small problems.  On the other hand, given
that the state of the art in quantum computing is largely limited by ten
qubits, such small circuits are of interest to physicists building quantum
computing devices.

\section{Quantum Search Applications}
\label{sec:grover}

Quantum computation is necessarily reversible, and quantum
circuits generalize their reversible counterparts in the classical
domain \cite{NielsenChuang}. Instead of wires, information is
stored on {\em qubits}, whose states we write as $\ket{0}$ and
$\ket{1}$ instead of $0$ and $1$. There is an added complexity ---
a qubit can be in a {\em superposition state} that combines
$\ket{0}$ and $\ket{1}$. Specifically, $\ket{0}$ and $\ket{1}$ are
thought of as vectors of the {\em computational basis}, and the
value of a qubit can be any unit vector in the space they span.
The scenario is similar when considering many qubits at once: the
possible configurations of the corresponding classical system
(bit-strings) are
now the computational basis, and any unit vector in the linear
space they span is a valid configuration of the quantum system.
Just as the classical configurations of the circuit persist as
basis vectors of the space of quantum configurations, so too
classical reversible gates persist in the quantum context.
Non-classical gates are allowed, in fact, any (invertible)
norm-preserving linear operator is allowed as a quantum gate.
However, quantum gate libraries often have very few non-classical
gates \cite{NielsenChuang}. An important example of a
non-classical gate (and the only one used in this paper) is the
Hadamard gate $H$. It operates on one qubit, and is defined as
follows: $H\ket{0}=\frac{1}{\sqrt{2}}(\ket{0}+\ket{1})$, and
$H\ket{1}=\frac{1}{\sqrt{2}}(\ket{0}-\ket{1})$. Note that because
$H$ is linear, giving the images of the computational basis
elements defines it completely.

During the course of a computation, the quantum state can be
any unit vector in the linear space spanned by the computational basis.
However, a serious limitation is imposed by quantum measurement,
performed after a quantum circuit is executed. A measurement
non-deterministically collapses the state onto some vector in a
basis corresponding to the measurement being performed. The
probabilities of outcomes depend on the measured state --- basis
vectors [nearly] orthogonal to the measured state are least likely
to appear as outcomes of measurement. If $H\ket{0}$ were measured
in the computational basis, it would be seen as $\ket{0}$ half the
time, and $\ket{1}$ the other half.

\begin{figure}[t]
  \begin{center}
  \vspace{-5cm}
    \includegraphics[width=14cm]{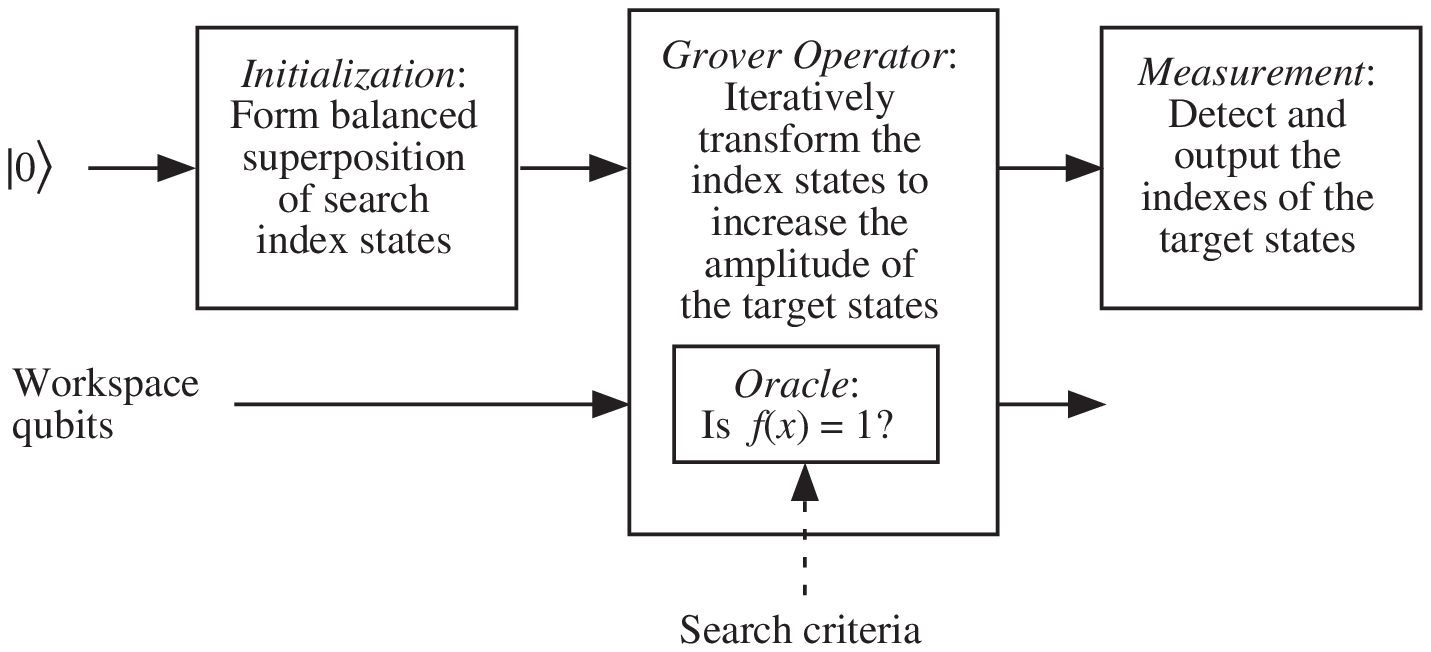}
  \caption{\label{fig:grover}
    A high-level schematic of Grover's search algorithm.
  }
 \vspace{2mm}
  \end{center}
\end{figure}

Despite this limitation, quantum circuits have significantly more
computational power than classical circuits. In this work, we
consider Grover's search algorithm, which is faster than any
known non-quantum algorithm for the same
problem \cite{Grover}. Figure \ref{fig:grover} outlines a possible
implementation of Grover's algorithm.  It begins by creating a
balanced superposition of $2^n$ n-qubit states which correspond to
the indexes of the items being searched. These index states are
then repeatedly transformed using a {\em Grover operator} circuit,
which incorporates the search criteria in the form of a
search-specific predicate $f(x)$. This circuit systematically
amplifies the search indexes that satisfy $f(x) = 1$ until a final
measurement identifies them with high probabliity.

\begin{figure}
  \begin{center}
    \includegraphics{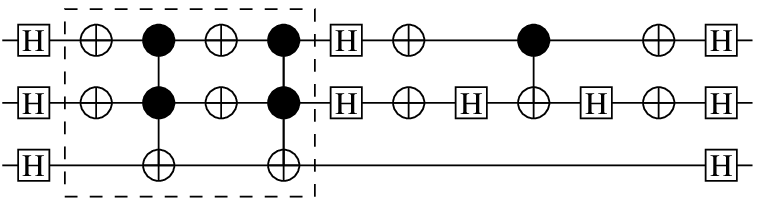}
  \caption{\label{fig:gropic}
    A Grover-operator circuit with oracle highlighted.
  }
 \vspace{2mm}
  \end{center}
\end{figure}

A key component  of the Grover operator is a so-called ``oracle''
circuit that implements a search-specific predicate $f(x)$. This
circuit transforms an arbitrary basis state $\ket{x}$ to the state
$(-1)^{f(x)}\ket{x}$. The oracle is followed by (i) several
Hadamard gates, (ii) a subcircuit which flips the sign on all
computational basis states other than $\ket{0}$, and (iii) more
Hadamard gates. A sample Grover-operator circuit for a search on 2
qubits is shown in Figure \ref{fig:gropic} and uses one qubit of
temporary storage \cite{NielsenChuang}. The search space here is
$\{0,1,2,3\}$, and the desired indices are $0$ and $3$.  The
oracle circuit is highlighted by a dashed line. While the portion
following the oracle is fixed, the oracle may vary depending on
the search criterion. Unfortunately, most works on Grover's
algorithm do not address the synthesis of oracle circuits and
their complexity. According to Bettelli et al.  \cite{Bettelli},
this is a major obstacle for automatic compilation of high-level
quantum programs, and little help is available.

\thm{LEM}{ \label{lem:plusOneQ} \cite{NielsenChuang}
  With one temporary storage
  qubit, the problem of synthesizing a quantum circuit that
  transforms computational basis states $\ket{x}$ to
  $(-1)^{f(x)}\ket{x}$ can be reduced to a problem in the synthesis
  of classical reversible circuits.
} {
  Define the
  permutation $\pi_f$ by $\pi_f(x,y)=(x,y \oplus f(x))$, and define
  a unitary operator $U_f$ by letting it permute the states of the
  computational basis according to $\pi_f$. The additional qubit is
  initialized to $\ket{-}=H\ket{1}$ so that $U_f
  \ket{x,-}=(-1)^{f(x)} \ket{x,-}$. If we now ignore the value of
  the last qubit, the system is in the state $(-1)^{f(x)} \ket{x}$,
  which is exactly the state needed for Grover's algorithm. Since a
  quantum operator is completely determined by its behavior on a
  given computational basis, any circuit implementing $\pi_f$
  implements $U_f$. As reversible gates may be implemented with
  quantum technology, we can synthesize $U_f$ as a reversible logic
  circuit.
}

Quantum computers implemented so far are severely limited by the
number of simultaneously available qubits. While $n$ qubits are
necessary for Grover's algorithm, one should try to minimize the
number of additional temporary storage qubits. One such qubit is
required by Lemma \ref{lem:plusOneQ} to allow classical reversible
circuits to alter the phase of quantum states.

\thm{COR}{
  For permutations $\pi_f(x,y)=(x,y \oplus f(x))$,
  such that $\{x:f(x)=1\}$ has even
  cardinality, no more temporary storage is necessary. For the
  remaining $\pi_f$, we need an additional qubit of temporary
  storage.
} {
  The permutation $\pi_f$ swaps $(x,y)$ with $(x,y \oplus f(x))$,
  and therefore performs one transposition for each element of
  $\{x:f(x)=1\}$. It is therefore even exactly when this set has
  even cardinality. The lemma follows from Corollary
  \ref{cor:theoddones}.
}

Given $\pi_f$, we can use the algorithm of Section \ref{sec:algos}
to construct an optimal circuit for it. Table \ref{tab:31oracdist}
gives the optimal circuit sizes of functions $\pi_f$ corresponding
to 3-input 1-output functions $f$ (``3+1 oracles'') which can be
synthesized on four wires. These circuits are significantly
smaller than many optimal circuits on four wires. This is not
surprising, as they perform less computation.

\begin{table}
  \small
  \begin{center}
    \begin{tabular}{|c||c|c|c|c|c|c|c|c|c|}\hline
      Circuit Size          & 0 & 1 & 2 & 3 &4 & 5& 6 & 7 & Total \\ \hline
      No. of circuits & 1 & 7 & 21& 35&35&24& 4 & 1 & 128\\ \hline
    \end{tabular}

  \caption{Optimal 3+1 oracle circuits for Grover's search.
           \label{tab:31oracdist}
           }
  \end{center}
\end{table}

In Grover oracle circuits, the main input lines preserve their
input values and only the temporary storage lines can change their
values.  Therefore, Travaglione et al. \cite{Travaglione} studied
circuits where some lines cannot be changed even at intermediate
stages of computation. In their terminology, a circuit with $k$
lines that we are allowed to modify and an arbitrary number of
read-only lines is called a $k$-bit {\em ROM-based circuit}. They show
how to compute permutation $\pi_f$ arising from a Boolean function
$f$ using a $1$-bit quantum ROM-based circuit, and prove that if
only classical gates are allowed, two writable bits are necessary.
Two bits are sufficient if the CNT gate library is used. The
synthesis algorithms of Travaglione et al. \cite{Travaglione}
rely on XOR sum-of-products decompositions of $f$.
We outline their method in a proof of the following result.

\thm{LEM}{ \label{lem:Trava}  \cite{Travaglione}
  There exists
  a reversible 2-bit ROM-based CNT-circuit computing
  $(x,a,b)\to (x,a,b \oplus f(x))$, where $x$ is a $k$-bit input.
  If a function's XOR decomposition
  consists of only one term, let $k$ be the number of literals appearing
  (without complementation).
  If $k>0$ then $3 \cdot 2^{k-1} -2$ gates are required.
} {
  Assume we are given an XOR sum-of-products decomposition of $f$.
  Then it suffices to know how to transform $(x,a,b) \to (x,a,b \oplus p)$
  for an arbitrary product of uncomplemented literals $p$,
  because then we can add the terms in an XOR
  decomposition term by term. So, without loss
  of generality, let $p=x_1 \ldots x_m$. Denote by
  $T(a,b;c)$ a T gate with controls on $a,b$ and inverter on $c$. Similarly,
  denote by $C(a;b)$ a C gate with control on $a$ and inverter on $b$.
  Number the ROM wires $1 \ldots k$, and the non-ROM wires $k+1$ and $k+2$.
  Let us first suppose that there is at least one uncomplemented literal,
  and put a $C(1;k+2)$ on the circuit; note that
  $C(1;k+2)$ applied to the input $(x,a,b)$ gives $(x,a,b\oplus x_1)$. We will
  write this as $C(1;k+2):(x,a,b)\to (x,a,b\oplus x_1)$, and denote this
  operation by $W_1$. Then, we define the circuit $W_2'$ as
  the sequence of gates $T(2,k+2;k+1) W_0 T(2,k+2;k+1) W_0$,
  and one can check that $W_2': (x,a,b) \to (x,a\oplus x_1 x_2,b)$.
  We define $W_2$ by exchanging the wires $k+1$ and $k+2$; clearly
  $W_2:(x,a,b)\to (x,a,b\oplus x_1 x_2)$.
  In general, given a circuit
  $W_l: (x,a,b\oplus x_1 \ldots x_{l-1}) \to (x,a \oplus x_1 \ldots x_l)$,
  we define $W_{l+1}' = T(l+1,k+2;k+1) W_l T(l+1,k+2;k+1) W_l$; one can check
  that $W_{l+1}':(x,a,b)\to (x,a\oplus x_1 \ldots x_{l+1},b)$. Define
  $W_{l+1}$ by exchanging the wires $k+1$ and $k+2$; then clearly
  $W_{l+1}:(x,a,b)\to (x,a,b\oplus x_1 \ldots x_{1+1})$. By induction,
  we can get as many uncomplemented literals in this product as we like.
}

The heuristic presented above has the property that none of its
gates has more than one control bit on a ROM bit. Indeed,
Travaglione et al. \cite{Travaglione} had restricted their
attention to circuits with precisely this property. However, they
note \cite{Travaglione} that their results do not depend on this
restriction.

We applied the construction of Lemma \ref{lem:Trava} to all 256
functions implementable in 2-bit ROM-based circuits with 3 bits of
ROM. The circuit size distribution is given in the line labeled
XOR in Table \ref{tab:3ROM2}. In comparing with circuits lengths
resulting from our synthesis algorithm of Section \ref{sec:algos},
we consider two cases. First, in the OPT T line, we only look at
circuits satisfying the restriction mentioned above. Then, in the
OPT line, we relax this restriction and give the circuit size
distribution for optimal circuits.\footnote{ Using a circuit
library with $\leq$ 6 gates (191Mb file, 1.5 min to generate), the
OPT line takes 5 min to generate. The use of a 5-gate library
improved the runtimes by at least 2x if we do not synthesize the
only circuit of size 11. For the OPT T line, we first find the 250
optimal circuits of size $\leq$ 12 (15 min) using a 6-gate library
(61Mb, 5min). The remaining 6 functions were synthesized in 5 min
with a 7-gate library (376Mb, 10 min). This required more than 1Gb
of RAM. }

\begin{table}
  \small
  \begin{center}
    \begin{tabular}{|l||l|l|l|l|l|l|l|l|l|l|l|l|l|l|l|l|l|l|l|l|l|l|l|l|l|l|l|}
      \hline
      Size & 0 & 1 & 2 & 3 & 4 & 5 & 6 & 7 & 8 & 9 & 10 & 11 & 12 & 13  \\
      \hline
      XOR & 1 & 4 & 6 & 4 & 4 & 12 & 18 & 12 & 6 & 12 & 19  & 16 & 10 & 8 \\
      \hline
      OPT T& 1 & 4 & 6 & 4 & 4 & 12 & 21 & 24 & 29 & 33 & 44 & 46 & 22 & 5\\
      \hline
      OPT & 1 & 7 & 21 & 35 & 36 & 28 & 28 & 36 & 35 & 21 & 7 & 1 & 0 & 0 \\
      \hline
    \end{tabular}

    \vspace{4mm}

    \begin{tabular}{|l||l|l|l|l|l|l|l|l|l|l|l|l|l|l|l|l|l|l|l|l|l|l|l|l|l|l|l|}
      \hline
      Size & 14 & 15 & 16 & 17 & 18 & 19 & 20 & 21 & 22 & 23 & 24 & 25 & 26 \\
      \hline
      XOR & 10 & 16 & 19 & 12 & 6 & 12 & 18 & 12  & 4  & 4  & 6  & 4  & 1 \\
      \hline
      OPT T & 1 & 0 & 0 & 0 & 0 & 0 & 0 & 0 & 0 & 0 & 0 & 0 & 0 \\
      \hline
      OPT & 0 & 0 & 0 & 0 & 0 & 0 & 0 & 0 & 0 & 0 & 0 & 0 & 0 \\
      \hline
    \end{tabular}
  \end{center}
  \caption{\label{tab:3ROM2}
    Circuit size distribution
    of 3+2 ROM-based circuits synthesized using various algorithms.}
\end{table}

Most functions computable by a 2-bit ROM-based circuit actually
require two writeable bits \cite{Travaglione}. Whether or not a
given function can be computed by a 1-bit ROM-based CNT-circuit,
can be determined by the following constructive procedure. Observe
that gates in 1-bit ROM circuits can be reordered arbitrarily, as
no gate affects the control bits of any other gate. Thus, whether
or not a C or T gate flips the controlled bit, depends only on the
circuit inputs. Furthermore, multiple copies of the same gate on
the same wires cancel out, and we can assume that at most one is
present in an optimal circuit. A synthesis procedure can then
check which gates are present by applying the permutation on every
possible input combination with zero, one, or two 1s in its binary
expansion. (Again, we have relaxed the restriction that only 1
control may be on a ROM wire). If the value of the function is
$1$, the circuit needs an N, C or T gate controlled by those bits.

Observe that adding the S gate to the gate library during $k+1$
ROM synthesis will never decrease circuit sizes --- no two wires
can be swapped since at least one of them is a ROM wire. In the
case of $k+2$ ROM synthesis, only the two non-ROM wires can be
swapped, and one of them must be returned to its initial value by
the end of the computation. We ran an experiment comparing circuit
lengths in the 3+2 ROM-based case and found no improvement in
circuit sizes upon adding the S gate, but we have been unable
to prove this in the general case.

\section{Conclusions} \label{sec:conclusion}

We have explored a number of promising techniques for synthesizing
optimal and near-optimal reversible circuits that require little
or no temporary storage.   In particular, we have proven
that every even permutation function can be
synthesized without temporary storage using the CNT gate library.
Similarly, {\em any} permutation, even or odd, can be
synthesized with up to one bit of temporary storage.
We have recently discovered that A. DeVos has independently demonstrated this
result, however, his proof relies on non-trivial
group-theoretic notions and resorts to a computer algebra package for a
special case.
\cite{DeVos}
We give a much more elementary analysis, and moreover our proof techniques 
are sufficiently constructive to be interpreted as a synthesis heuristic. 
We have also derived various equivalences among CNT-circuits that are useful
for synthesis purposes, and given a decomposition of a CNT-circuit into a
T$|$C$|$T$|$N-circuit. 

To further investigate the structure of reversible circuits, we
developed a method for synthesizing optimal reversible circuits.
 While this algorithm scales
better than its counterparts for irreversible computation
\cite{Lawler}, its runtime is still exponential. Nonetheless, it
can be used to study small problems in detail, which may be of
interest to physicists building quantum computing devices because
the current state of the art is largely limited by 10 qubits. One
might think that an exhaustive search procedure would suffice for
small problems, but in fact, even for three-input circuits, an
exhaustive search is nowhere near finished after 15 hours; our
procedure terminates in minutes. Our experimental data about {\em
all} optimal reversible circuits on three wires using various
subsets of the CNTS library reveal some interesting
characteristics of optimal reversible circuits. Such statistics,
extrapolated to larger circuits, can be used in the future to
guide heuristics, and may suggest new theorems about reversible
circuits.

Finally, we have applied our optimal synthesis tool to the design of
oracle circuits for a key quantum computing application, Grover's
search algorithm, and obtained much smaller circuits than previous
methods. Ultimately, we aim
to extend the proposed methods to handle larger and more general
circuits, with the eventual goal of synthesizing quantum circuits
containing dozens of qubits.

\newpage

\addtolength{\baselineskip}{-0.7mm}

\end{document}